\documentclass[twocolumn,superscriptaddress,showpacs,prd,aps,amsmath,amssymb,nofootinbib]{revtex4-1}

\usepackage{graphicx}
\usepackage{longtable}

\newcommand{\beq}{\begin{equation}} 
\newcommand{\eeq}{\end{equation}} 
\newcommand{\beqn}{\begin{eqnarray}} 
\newcommand{\eeqn}{\end{eqnarray}}

\newcommand{\zD}{{\raise1.0ex\hbox{${}^{\ \circ}$}}\!\!\!\!\!D}
\newcommand{\alone}{{\raise0.5ex\hbox{${}^{\ 1}$}}\!\!\!\!\alpha}

\newcommand{\nalam}{\mathrel{\raise0.9ex\hbox{$^\lambda$}\mkern-14mu
\lower0.0ex\hbox{$\nabla$}}}

\newcommand{\zeroD}{{\raise1.0ex\hbox{${}^{\ \circ}$}}\!\!\!\!\!D}
\newcommand{\zLap}{{\raise1.0ex\hbox{${}^{\ \circ}$}}\!\!\!\!\Delta}
\newcommand{\zna}{{\raise1.0ex\hbox{${}^{\ \circ}$}}\!\!\!\!\!\nabla}
\newcommand{\zS}{{\raise1.0ex\hbox{${}^{\ \circ}$}}\!\!\!\!\!S}

\usepackage[normalem]{ulem}
\usepackage{epstopdf}
\usepackage{color}
\usepackage{soul}
\usepackage{bm}
\usepackage{xspace}

\newcommand{\cocal}{\textsc{cocal}\xspace}

\usepackage[]{amsmath}
\usepackage[]{amsfonts}
\usepackage[]{amssymb}
\usepackage[]{mathrsfs}
\usepackage[]{mathtools}
\usepackage{times}
\usepackage{hyperref}
\hypersetup{
  colorlinks=true,        
  linkcolor=blue,         
  citecolor=cyan,         
}
\def\QEQ{{%
			\setbox0\hbox{$I$}%
			\rlap{\hbox to \wd0{\hss--\hss}}\box0
		}}

\begin{document}

\title{Differentially rotating strange star in general relativity}

\author{Enping Zhou}
\affiliation{Max Planck Institute for Gravitational Physics (Albert Einstein
Institute), Am M\"uhlenberg 1, Potsdam-Golm, 14476, Germany} \affiliation{State
Key Laboratory of Nuclear Science and Technology and School of Physics, Peking
University, Beijing 100871, People's Republic of China} 

\author{Antonios Tsokaros}
\affiliation{Department of Physics, University of Illinois at Urbana-Champaign,
Urbana, IL 61801, USA}

\author{K\=oji Ury\=u}
\affiliation{Department of Physics, University of the Ryukyus, Senbaru,
Nishihara, Okinawa 903-0213, Japan}

\author{Renxin Xu}
\affiliation{State Key Laboratory of Nuclear Science and Technology and
  School of Physics, Peking University, Beijing 100871, People's Republic
  of China} \affiliation{Kavli Institute for Astronomy and Astrophysics,
  Peking University, Beijing, 100871, People's Republic of China}

\author{Masaru Shibata}
\affiliation{Max Planck Institute for Gravitational Physics (Albert Einstein
Institute), Am M\"uhlenberg 1, Potsdam-Golm, 14476, Germany} \affiliation{Center
for Gravitational Physics, Yukawa Institute for Theoretical Physics, Kyoto
University, Kyoto, 606-8502, Japan}

\date{\today}

\begin{abstract}
Rapidly and differentially rotating compact stars are believed to be formed in
binary neutron star merger events, according to both numerical simulations and
the multi-messenger observation of GW170817.  Questions that have not been
answered by the observation of GW170817 and remain open are whether or not a
phase transition of strong interaction could happen during a binary neutron star
merger event that forms a differentially rotating strange star as a remnant, as
well as the possibility of having a binary strange star merger scenario. The
lifetime and evolution of such a differentially rotating star, is tightly
related to the observations in the post-merger phase. Various studies on the
maximum mass of differentially rotating neutron stars have been done in the
past, most of which assume the so-called $j$-const law as the rotation profile
inside the star and consider only neutron star equations of state. In this
paper, we extend the studies to strange star models, as well as to a new rotation
profile model.  Significant differences are found between differentially
rotating strange stars and neutron stars, with both the $j$-const law and the new
rotation profile model. A moderate differential rotation rate for neutron stars
is found to be too large for strange stars, resulting in a rapid drop in the
maximum mass as the differential rotation degree is increased further from
$\hat{A}\sim2.0$, where $\hat{A}$ is a parameter characterizing the differential 
rotation rate for $j$-const law. As a result the maximum mass of a differentially rotating
self-bound star drops \textit{below the uniformly rotating mass shedding limit}
for a reasonable degree of differential rotation. 
The continuous
transition to the toroidal sequence is also found to happen at a much smaller
differential rotation rate and angular momentum than for neutron
stars. In spite of those differences, $\hat{A}$-insensitive relation between the maximum mass
for a given angular momentum is still found to hold, even for the new differential rotation law.
Astrophysical consequences of these differences and how to distinguish
between strange star and neutron star models with future observations are also
discussed.

\end{abstract}

\maketitle

\section{Introduction}
In the coming multi-messenger astronomy era led by the observation of GW170817
\citep{Abbott2017_etal} and its electromagnetic (EM) counterparts
\citep{Abbott2017b}, it's very likely that a conclusion could be drawn on the
equation of state (EoS) of compact stars, which is a challenging topic in
nuclear physics due to the non-perturbative nature of strong interaction at low
energy scale. In fact, GW170817 alone has already provided ample information on
the radius of neutron stars (NSs) by measuring the tidal deformability in the
gravitational wave (GW) signal at the late inspiral stage (c.f. systematic
studies in \citep{Annala2017,Abbott2018tidal}).  Moreover, constraints on the
maximum mass have also been put forward by considering the fate of the merger
remnant together with the electromagnetic counterparts of GW170817
\citep{Rezzolla2017,Ruiz2017,Shibata2017c,Bauswein2017b}.

However, besides conventional NS EoSs, other possibilities such as stars composed of 
strange quark matter \citep{Farhi1984,Alcock86,Haensel1986}, namely strange star (SS) models, 
are not excluded by the observation of GW170817
\citep{Abbott2017_etal}.  In addition, the EM counterparts of
GW170817 could also be understood within the scenario of a binary strange star
(BSS) merger \citep{Li2016,Lai2018,Bauswein2009,Paulucci2017}. Because of their
self-bound nature, SSs are quite different from NSs. The tidal deformability
measurement from GW170817 will imply a different radius constraint if the SS
branch is taken into account \citep{Monata2018,Most2018}. For the case that it is
supported by rigid rotation, the maximum mass of SSs can be increased much more 
than NSs \citep{Li2016}. Rotating SSs can reach much higher $T/|W|$ ratio than NSs,
leading to a more important role of triaxial instabilities for the case that the rotation is
fast enough \citep{Zhou2018,rosinska2000b}. Even in the case of a binary neutron
star (BNS) merger, whether or not a phase transition and the formation of a SS
happens during the merger will significantly alter the GW signals
\citep{Most2018b}.  Considering all of the above, it's also quite important to
calculate models of differentially rotating strange stars, which has never been
done before, to better understand the observation of binary merger events.

Depending on the maximum mass of the EoS and the total mass of the merging
binary, there could be several different outcomes after the merger: a prompt
collapse to a black hole, a short-lived hypermassive neutron star (HMNS, the
mass of which exceeds the mass-shedding limit with rigid rotation, hence is only
stable with differential rotation) or a long-lived supramassive neutron star.
The amount and the velocity of the ejected mass in the post-merger phase, the
neutrino emission as well as the energy injection from the merger remnant is
quite different in every case. Therefore, it's possible to make constraints on
the remnant type, hence the maximum mass of the EoS, according to the EM
counterparts of the merger event. 


Following the evolution of a differentially rotating compact star in the
post-merger phase for a long time is computationally expensive. Therefore the
study of equlibrium models is very useful, especially when one is concerned with
the parameter space explorations (e.g.  \citep{Breu2016,Bozzola2018}). Also the
evolution of SSs is a numerically challenging problem due to its finite surface
density.  As a result, calculating differentially rotating SSs is an effective
way to study the outcome of merger events for the hypothetical SS formation. The
choice of a differential rotation law (i.e., the angular velocity as a function
of the cylindrical radial coordinate $\Omega=\Omega(r\sin\theta)$ in the
Newtonian case) is essential for modeling differentially rotating stars. In the
case of relativistic gravity, instead, one has to choose the relativistic
specific angular momentum as a function of angular velocity (i.e.,
$j=j(\Omega)$, in which $j\coloneqq u^t u_\phi$ and $u^\alpha$ is the 4-velocity
of the fluid).  The most commonly used differential rotation law is the
so-called $j$-const law
\citep{Komatsu89,Ansorg2009,Rosinska2017,Studzinska2016,Baumgarte00b,Morrison2004,Kaplan2014},
\beq
j(\Omega)=A^2(\Omega_\mathrm{c}-\Omega), 
\label{eq:jconst} 
\eeq 
in which $A$ and $\Omega_\mathrm{c}$ are two constant parameters in the model.
A dimensionless parameter $\hat{A}=A/r_{\rm e}$ is also quite often used, where $r_{\rm e}$
is the equatorial radius of the star.
This choice results in a monotonically decreasing angular velocity with respect
to the cylindrical radius. However, it has been realized that such a
differential rotation profile is not realistic from numerical simulations of
BNSs mergers. In the equatorial plane, simulations suggest that the angular
velocity starts from a nonzero finite value on the rotational axis; then
increases towards a maximum value; and then decreases to a minimum 
\citep{Shibata05c,Hotokezaka2013c,Dietrich2015,Bauswein2015,Kastaun2014,Hanauske2016}.
Hence, it's quite interesting and important to model differentially rotating
stars with such a rotation law, as is done in \citep{Uryu2017}.

In this paper, we have applied both the $j$-const law as well as a more
realistic rotation law to SS models. The Compact Object CALculator (\cocal) code
which we have modified to include self-bound stars and tested its convergence
and accuracy before \citep{Zhou2018}, is used for constructing the equilibrium
solutions. We have compared our results to those of neutron stars and found that
for differentially rotating SSs, both the drop of the maximum mass and the
transition to the toroidal sequence happens at much larger differential rotation
rate, compared with the results of NSs. Interestingly enough, the maximum
mass of a differentially rotating SS can be smaller than that of a rigidly
rotating one for both differential rotation laws with a reasonable differential
rotation rate.

The paper is organized as follows: the SS EoSs used in this paper
will be introduced in Sec.\ref{sec:eos}. In Sec \ref{sec:models} we briefly
review the formulations and differential rotation laws used in the calculation.
The results will be presented in Sec. \ref{sec:results}. The astrophysical
implications of those results will be discussed in Sec. \ref{sec:disandconclu}.
Note that in this paper we use units with $G=c=M_\odot=1$ unless otherwise stated. 
Here $G$ and $c$ are the gravitational constant and speed of light, respectively.

\section{Strange Star Equation of States}
\label{sec:eos}

In this work, we have considered two types of EoS for SSs.  One of them is the
widely used MIT bag model \cite{chodos1974}. As we are only interested in the
self-bound nature of SSs and its impact of differential rotation,
the effects of perturbative quantum choromodynamics (QCD) due to gluon mediated
quark interactions \citep{Fraga2001,Alford2005,Li2017,Bhattacharyya2016} will
not be considered, nor the finite mass of the strange quark.  This allows us to
have a much simpler EoS model for numerical calculations (similar to e.g.
\citep{limousin2005}), in which pressure is related to \textit{total energy
density} according to
\beq
p=1/3 (\epsilon-\epsilon_s) \, ,
\label{eq:mit_gen}
\eeq
where $\epsilon_s=4B$ is the total energy density at the surface and $B$ the bag
constant \citep{Alcock86,Haensel1986}. $p$ and $\epsilon$ are pressure and total energy
density of the matter, respectively. In this work,  $B$ is chosen to be
$(138\,\mathrm{MeV})^4$.

Another EoS model considered in this work is the so-called strangeon star model
\citep{Lai2017}.  Unlike the MIT bag model in which quarks are assumed to be
de-confined and described by Fermi gas approximation, Lai and Xu suggested that
clustering of quarks is possible at the density of a cold compact star since the
coupling of strong interaction is not negligible at such energy scale. Lai and
Xu attempted to approach the EoS with phenomenological models, i.e., to compare
the potential with the interaction between inert molecules \citep{lai2009} (a
similar approach has also been discussed in \cite{Guo2014}). They also take the
lattice effects into account as the potential could be deep enough to trap the
strangeons. Combining the inter-cluster potential and the lattice
thermodynamics, an EoS could be derived in terms of number density of constituent strangeon ($n$):
\beq
p=4U_0(12.4r_0^{12}n^5-8.4r_0^6n^3)+\frac{1}{8}(6\pi^2)^{\frac{1}{3}}\hbar cn^{\frac{4}{3}}.
\label{eq:lx09eos}
\eeq
The parameters $U_0$ and $r_0$, are the depth of the potential and the
characteristic range of the interaction, respectively\footnote{Note that this
equation has a unique non-zero root, demonstrating the self-bound nature of
strangeon star model.}. The EoS depends also on the number of quarks in each
strangeon particle ($N_\mathrm{q}$). Similar to the MIT bag model case, we use
the rest-mass density parameter in the numerical code, which is 
\beq
\rho=m_\mathrm{u}\frac{N_\mathrm{q}}{3}n,
\eeq
where $m_\mathrm{u}=931\mathrm{MeV}/c^2$ is the atomic mass unit. In this work
the model with $U_0=50\,\mathrm{MeV}$ and $N_\mathrm{q}=18$ is chosen.
The details about the explicit
implementation of SS models in the \cocal code are explained in detail in our
previous work \citep{Zhou2018}.

Both the MIT bag model and the strangeon star model used in this work satisfy
the maximum mass constraint by the discovery of massive pulsars
\cite{Demorest2010,Antoniadis2013,Zhou2018} as well as the tidal deformability
constraint by GW170817 (\citep{Abbott2017_etal,Zhou2018b,Lai2018}, also c.f.
Table.\ref{tab:tovmax}). It's worth to
remark that there is a positive correlation between the maximum mass and tidal
deformabilty for NS EoSs as they both relate to the stiffness of the EoS model.
According to Fig.1 in \citep{Annala2017}, in order to satisfy the tidal
deformability constraint, there will be an upper limit for the maximum mass of
any NS EoSs. This correlation holds qualitatively for SSs (c.f.
\citep{Lai2018b,Zhou2018b}) but not quantitatively due to the finite surface
density of SSs which leads to a correction in the calculation of tidal
deformablity \citep{Damour:2009,Postnikov2010}. As a result, it's much easier
for strange star models to accommodate the observation of GW170817 and massive
pulsars at the same time.  Additionally, previous studies have also demonstrated
the possibility of understanding some puzzling observations within SS scenario,
such as the energy release during pulsar glitches \citep{zhou2014}, the peculiar
X-ray flares \citep{xu2009}, the optical/UV excess of X-ray-dim isolated neutron
stars \cite{wang2017} as well as the multiple internal plateau stages in short
gamma bursts \citep{Hou2018}.

\begin{table}
	\begin{tabular}{cccccccccccc}
		\hline
		EOS & $\rho_\mathrm{surf}$ & $M_\mathrm{TOV}$ & $\rho_\mathrm{c,TOV}$ & $R_{1.4}\,[\mathrm{km}]$ & $\Lambda_{1.4}$ \\
		\hline
		MIT & $1.4\rho_0$ &	2.217 & $5.42\rho_0$ & 11.814 & 792.8	 \\
		LX  & $2\rho_0$   & 3.325 & $4.03\rho_0$ & 10.459 & 381.9	 \\
		\hline
	\end{tabular}
	\caption{Surface density ($\rho_\mathrm{surf}$), TOV maximum mass
($M_\mathrm{TOV}$), central density for the TOV maximum mass solution
($\rho_\mathrm{c,TOV}$) for the two EOSs in this work. The densities are in
units
of nuclear saturation density ($\rho_0=2.67\times10^{14}\,\mathrm{g\,cm^{-3}}$).
We also show the radius and tidal deformability for a 1.4 solar mass star
for both EoSs.}
	\label{tab:tovmax}
\end{table}

In Table.\ref{tab:tovmax}, we list some properties of the two EoS considered in
this work. The MIT bag model has a much larger ratio between central density and
surface density compared with the strangeon star model for the
Tolman-Oppenheimer-Volkoff (TOV) maximum mass solution (i.e., 5.42/1.4 versus
4.03/2). This result indicates that the strangeon star model is more similar to
an incompressible EoS than MIT bag model quantitatively.  Moreover,
this difference in incompressibility will remain the same regardless of the bag
constant we are using for MIT model.  As pointed out by
\citep{Haensel1986,rosinska2000b}, when neglecting strange quark mass and
interaction between quarks mediated by gluons (as the model used in this paper),
the properties of the maximum mass solution for both rotating and non-rotating
cases simply rescale with the bag constant, keeping
$\rho_\mathrm{c}/\rho_\mathrm{surf}$ unchanged. This quantitative difference
between the two models will be discussed again in Sec.\ref{sec:mmax}.

\section{Differential rotation models}
\label{sec:models}

The hydrostatic equation in equilibrium can be derived from the
conservation of energy-momentum, $\nabla_\mu T^{\mu\nu}=0$, in which
$T^{\mu\nu}=(\epsilon+p)u^\mu u^\nu+pg^{\mu\nu}$ is the energy-momentum tensor
of a perfect fluid. For stationary and axisymmetric differential rotating stars
the Euler equation becomes \cite{Uryu2016a}
\beq
\nabla_\mu \ln{\frac{h}{u^t}} + u^t u_\phi \nabla_\mu \Omega - \frac{T}{h}\nabla_\mu s =0,
\label{eq:relahydro}
\eeq
where $h=(\epsilon+p)/\rho$ is the specific enthalpy, $\rho$ the rest mass
density, $T$ the temperature, and $s$ the specific entropy. Assuming isentropic
configurations, Eq. (\ref{eq:relahydro}) can be integrated as 
\beq
\frac{h}{u^t}\exp[{\int j d\Omega}]=\mathcal{E},
\label{eq:eulerint}
\eeq
provided an integrability condition $j\coloneqq u^t u_\phi=j(\Omega)$ is assumed.
$\mathcal{E}$ in Eq. (\ref{eq:eulerint}) is a constant to be determined
once the axis ratio and central density of the star is fixed.

The choice of a differential rotation law is exactly a choice for $j(\Omega)$.
As explained in \citep{Uryu2016a}, a simple generalization of the $j$-const law
(Eq. (\ref{eq:jconst})) is 
\beq
j(\Omega)=A^2 \Omega[(\frac{\Omega_\mathrm{c}}{\Omega})^q-1],
\label{eq:jconstcocal}
\eeq
where $\hat{A}$ is a parameter characterizing the differential rotation rate,
$\Omega_{\rm c}$ is the angular velocity along the rotation axis and $q$ is a new parameter.
Setting $q=1$ one recovers the $j$-const law. 
In the \cocal code normalized coordinates are used 
(equatorial radius of the star is normalized to 1), thus parameter $A$ in
Eq. (\ref{eq:jconstcocal}) is the same as $\hat{A}$ in other studies such
as \citep{Baumgarte00b}. The rotation profile reduces to
rigid rotation in the limit of $A\rightarrow\infty$.

Apart from the $j$-const law, we have also considered a more realistic
differential rotation profile used in \citep{Uryu2017} which mimics the
nonmonotonic $\Omega$ distribution as observed in the HMNS remnant formed in BNS
simulations \cite{Hanauske2016,Kastaun2014}. It should be reminded that for such
a nonmonotonic differential rotation profile $j(\Omega)$ becomes a multi-valued
function. Hence the integrability condition is written as $\Omega=\Omega(j)$
instead. As described in \citep{Uryu2017} we use 
\beq
\Omega=\Omega_\mathrm{c}\frac{1+(j/B^2\Omega_\mathrm{c})^p}{1+(j/A^2\Omega_\mathrm{c})^{q+p}}\, ,
\label{eq:oj}
\eeq
where $A,\ B,\ p$, and $q$ are parameters that control the
differential rotation profile. For the integration in Eq. (\ref{eq:eulerint}),
the following rearrangement is applied
\beq
\int jd\Omega=\int j\frac{d\Omega}{dj}dj.
\label{eq:ojint}
\eeq
The choice of $(p,q)$ is (1,3) in our calculations. For this law, 
rather than fixing $A$ and $B$, we choose to fix the
ratio between the maximum angular velocity and the central angular velocity
($\Omega_\mathrm{m}/\Omega_\mathrm{c}$) as well as the equatorial angular
velocity with respect to the central ($\Omega_\mathrm{eq}/\Omega_\mathrm{c}$)
and then solve for the corresponding $A$ and $B$ iteratively for each solution
\citep{Uryu2017}. 

Fixing the two angular velocity ratios mentioned above, we find that the
corresponding $A$ and $B$ parameters vary more significantly for SSs with
different central densities and axis ratios, than in NSs \citep{Uryu2017}.  For
solutions with large central densities or close to the mass shedding limit this
affects the convergence of the method in a very delicate way. Hence, similar to
what is done in \citep{Uryu2017}, we concentrate on differential solutions with
several constant axis ratios (i.e., $R_\mathrm{z}/R_\mathrm{x}=0.25,0.5$ and
0.75) instead of exploring the entire parameter space. The results will be
demonstrated in the next section.

For the equations of the gravitational field we employ the
Isenberg-Wilson-Mathews (IWM) formulation \citep{Isenberg1980} which assumes the
spatially conformal flat approximation \cite{Zhou2018}. Its validity and
accuracy in calculating both rigidly rotating and differentially rotating
relativistic stars has been verified in \citep{Cook1996,Iosif2014}. According to 
our comparison as well as previous results, it will be useful to keep in mind that 
the quantities calculated and reported in this paper might have up to 2\% error for
global quantities (e.g. ADM mass) and up to 5\% error for local quantities (e.g. angular velocity).

\section{Results}
\label{sec:results}

In this section we present results for differentially rotating SSs both with
the $j$-const as well as with the more realistic law Eq. (\ref{eq:oj}). We focus
on the properties of the maximum mass and the transition to toroidal topologies
for the EoSs mentioned in Sec. \ref{sec:eos}.

\subsection{Maximum mass of differentially rotating strange star}
\label{sec:mmax}

Differentially rotating NSs could normally reach much higher maximum mass
compared with uniformly rotating ones, thus called HMNS. According to previous
investigations with both $\Gamma=2$ polytropic EoS \citep{Baumgarte00b} or more
realistic EoSs \citep{Weih2017,Bozzola2018}, the maximum mass of HMNS increases
as the differential rotation increases as long as the rotational profile is not
extremely differential. To be precise, $M_\mathrm{max}$ increases as $\hat{A}$
decreases for various NS EoSs for $\hat{A}\in(\sim1,\infty)$. For the case that
$\hat{A}$ is smaller than 1, the maximum mass can actually drop, although it is 
still larger than the uniformly rotating mass shedding limit\footnote{Neutron stars 
supported only by rigid rotation are called supramassive (SMNS) \citep{Cook92b}.} (c.f.
\citep{Rezzolla_book:2013}). 
The maximum possible mass of a differentially rotating model could be as high as
twice of the non-rotating maximum mass $M_\mathrm{TOV}$ \citep{Baumgarte00b} or 
even 2.5 times depending on the EoS models \citep{Espino:2019ebx}.

\begin{figure}
	\begin{center}
		\includegraphics[width=0.95\columnwidth]{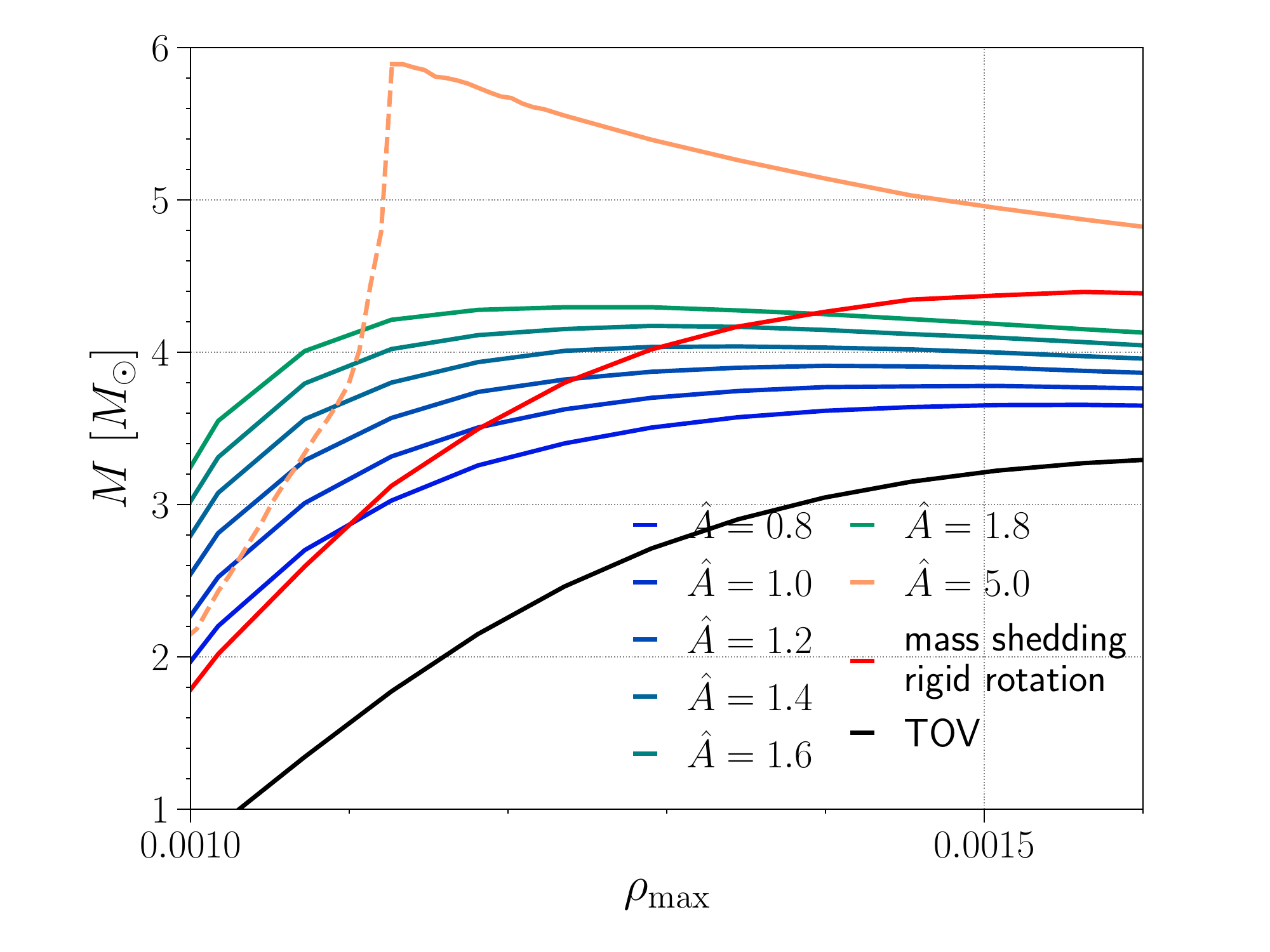}
	\end{center}
	\caption{Mass versus maximum density diagram for strangeon star model.
The black curve is for the non-rotating case (TOV solution) while the red
curve is for the mass shedding limit for uniformly rotating axisymmetric case.
Curves with gradually changing color from green to blue represents the mass
shedding limit for differentially rotating case with $j$-const law. The
$\hat{A}$ parameter applied to those curves range from 1.8 to 0.8 as the color
change from green to blue (from top to the bottom). Note that due to the
existence of type C solutions mentioned in Sec.\ref{sec:typec}, the maximum mass
of differentially rotating case could probably be found for the case that the central density
is not the maximum density inside star. We have also shown the $\hat{A}=5$ case which corresponds
to the maximum possible mass in our calculation in the yellow curve on top. For this particular differential
rotation rate, type C solutions and toroidal limit are only found for large central densities (i.e.,
$\rho_{\rm max}>1.12\times10^{-3}$). We label the part where only type A solutions exist by dashed curve. } 
\label{fig:lxmrhoc}
\end{figure}

Regarding the maximum mass of differentially rotating star, it's important to clarify 
the configuration types. As first pointed out by \citep{Ansorg2009}, there are 4 types
of differentially rotating neutron stars.
For small differential rotation rate, differentially rotating
star has a mass shedding limit when the star is still ellipsoidal (type A).
Whereas for moderate differential rotation rates, there exists type B and C solutions, for 
which the maximum mass is at the toroidal limit ($R_z/R_x=0$). The difference between type B and C
is that the later can smoothly transit into an ellipsoidal sequence and eventually a spherical star by
reducing angular momentum whereas the former one cannot and terminates at $R_z/R_x<1$ when losing 
angular momentum. Note that when the differential rotation rate is modest, there is also type D solutions
co-exists with type C solution (as type B co-exists with type A), which have two mass shedding limit 
but no toroidal or spherical limit.
For SSs, we have found that only type C solutions exist for most of the 
$\hat{A}$ parameter range we considered. In another word, type A and B solutions vanishes at 
much smaller differential rotation rate for SSs compared with NSs. Details will be explained again in Section.\ref{sec:typec}. There is indeed one model we have shown in Fig.\ref{fig:lxmrhoc}, which 
type A and C solutions still co-exist at different maximum density range for $\hat{A}=5.0$ and we are 
showing the maximum mass of them respectively (in dash and solid). For all the other cases, without further
mention, the maximum mass case is for type C configuration. 


In order to investigate the maximum mass of a hypermassive strange star (HMSS) and
its dependence on the $\hat{A}$ parameter, we have calculated HMSS models with
the $j$-const law and various choices of $\hat{A}$ ranging from 0.6 to 6.  This
will enable us to make a direct comparison with the HMNS models which obey the same
differential rotating law. Solutions are calculated for both the strangeon star
model and MIT bag model mentioned above.

\begin{figure}
	\begin{center}
		\includegraphics[width=0.95\columnwidth]{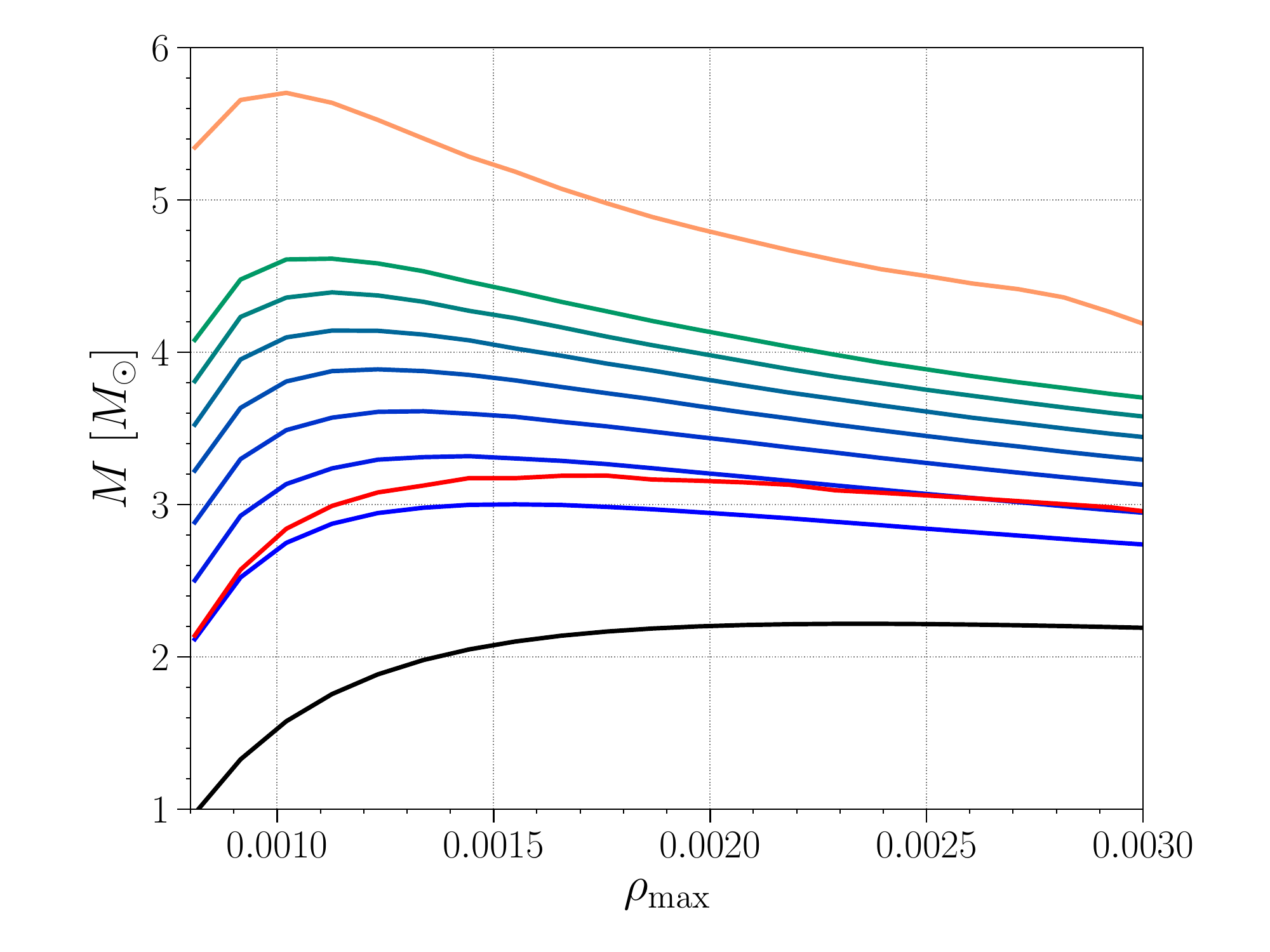}
	\end{center}
	\caption{Mass versus maximum density diagram for MIT bag model SSs. 
The models for curves with different colors are exactly the same as in
Fig.\ref{fig:lxmrhoc}. We calculated one more model for MIT bag model with $\hat{A}=0.6$
as shown by the bottom blue curve. The yellow curve on the top which corresponds to the maximum
possible mass case for MIT bag model is with $\hat{A}=3.0$.}
	\label{fig:mitmrhoc}
\end{figure}

The broadbrush picture of HMSSs with $j$-const law is similar to that of HMNSs,
but the quantitative dependence on the $\hat{A}$ parameter (namely the
differential rotation rate) is quite different from what was mentioned in the
paragraphs above. As $\hat{A}$ parameter approaches infinity, the rigid
rotation mass shedding limit will be recovered for HMSSs. Decreasing
$\hat{A}$ from infinity results to an increase of the maximum mass of HMSSs, up
until $\hat{A}\sim5$ for both strangeon star model and $\sim3$ for MIT bag model
(the corresponding value for HMNSs is around 1). This maximum possible mass for HMSSs 
is above 5 $M_\odot$. As $\hat{A}$ is further decreased from $\hat{A}\sim 5$, the maximum
mass begins to decrease (as in the HMNS case). We have chosen several models
with $\hat{A}$ ranging from 2 to 0.5 in Fig. \ref{fig:lxmrhoc} and Fig.
\ref{fig:mitmrhoc} to better illustrate the difference compared with HMNSs with
a moderate differential rotation rate.

There are several interesting points in the results shown in the
figures. First of all, as pointed by \cite{rosinska2000b}, the maximum mass of a
rigidly rotating SS (red curve) is approximately 40$\%$ larger than
$M_\mathrm{TOV}$ (black curve) for both EoSs, almost twice as large as the case
of NS EoSs \citep{Breu2016}.  Secondly, compared with the results of polytropic
NSs with $\Gamma=2$ shown in Fig. 1 in \cite{Baumgarte00b} where the maximum
mass of HMNSs increases significantly from $\hat{A}=2.0$ to 1.0, the maximum
mass of HMSSs actually decreases significantly in the exactly same range of
$\hat{A}$. In other words, while $\hat{A}=2.0$ is a small differential rotation
degree for NSs, it corresponds to very large one for SSs.  This is
understandable, considering the self-bound nature of SSs. SSs have finite
surface densities which are of the same order of magnitude as the central
density. In this sense, SSs are more like an incompressible star. In the case of
NSs, varying the equatorial angular velocity has a smaller effect since the
density at the equator approaches zero.  For SSs the situation is completely
different, and the configuration of the star is affected much more
by differential rotation.

\begin{figure}
	\begin{center}
		\includegraphics[width=0.95\columnwidth]{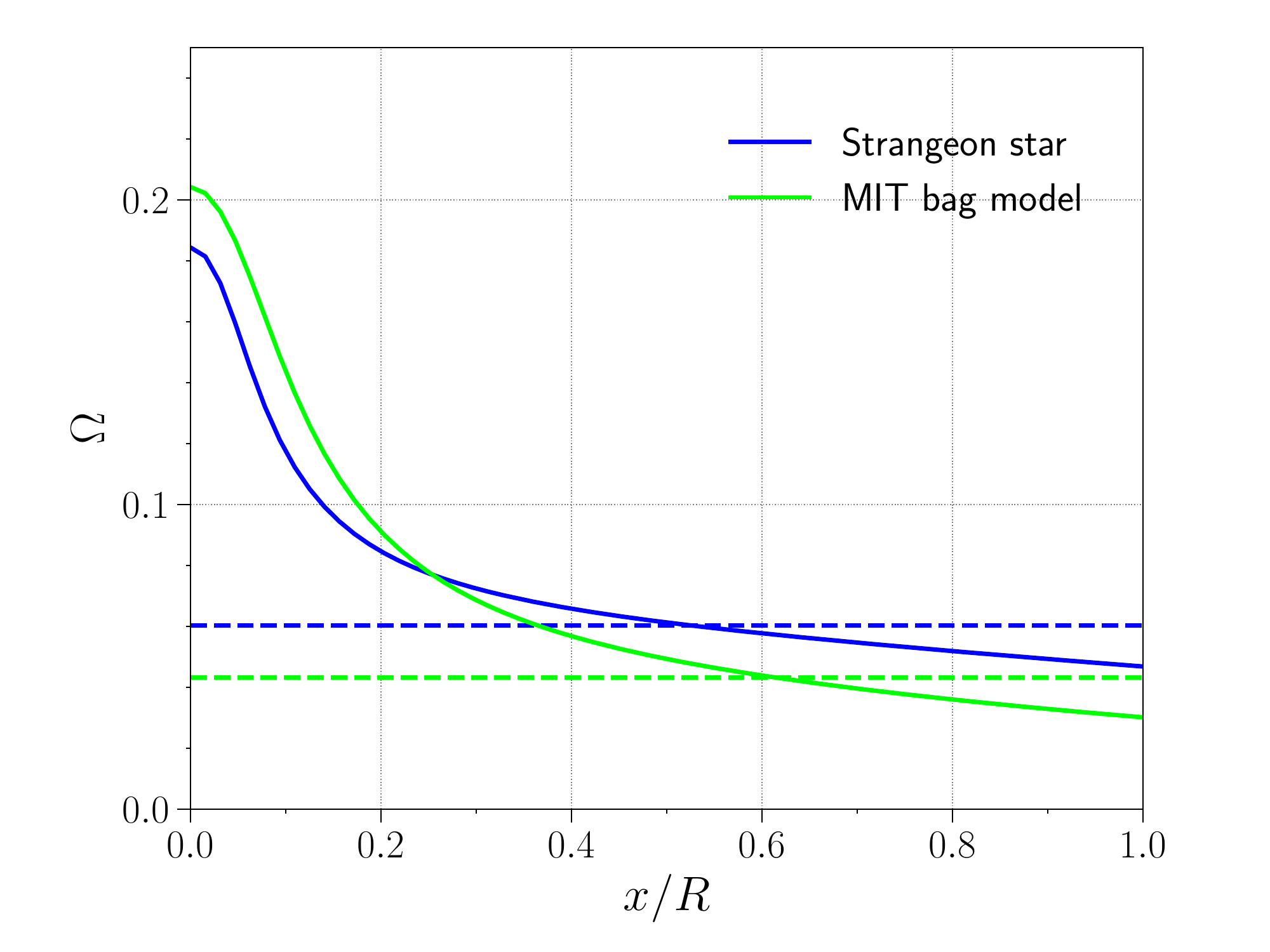}
	\end{center}
	\caption{The angular velocity profile for strangeon star (blue) and MIT bag model (green) 
	when the maximum mass becomes close to their rigidly rotating mass shedding limit. This means
    $\hat{A}=1.8$ for strangeon star model and $\hat{A}=0.8$ for MIT bag model. Dashed horizontal line 
    indicate the angular velocity for the mass shedding limit of rigid rotation case.}
	\label{fig:ome_max}
\end{figure}

Another interesting feature is that HMSSs can have a smaller maximum mass 
than in the rigid rotation case with a moderate differential rotation rate.
\footnote{ Note that this can in principle also happen for NSs, but with unrealistically extreme
differential rotation profile, e.g. $\hat{A}\sim0.1$ for a $\Gamma=2$ polytropic EoS.}
For the strangeon star this happens at $\hat{A}\sim1.8$ while for the MIT bag model 
at $\hat{A}\sim0.7$. 
Two aspects can account for this very interesting result: on one hand, due to the
finite surface density and larger incompressibility,
the maximum mass of strange stars drops more rapidly as
differential rotation is enhanced in strange stars; on the other hand, the supra-massive
mass shedding limit for SSs are much larger than NSs given the
same $M_\mathrm{TOV}$, making it possible for the HMSS maximum mass to drop
below it with moderate $\hat{A}$. The quantitative difference
for MIT bag model and strangeon star model could then also be interpreted by the difference in their
incompressibility, as mentioned in Sec.\ref{sec:eos}. In addition, the rotational profile
for the critical case where the maximum mass becomes comparable to mass shedding limit of the rigid rotation case can also be seen in Fig.\ref{fig:ome_max}. MIT bag model indeed needs a larger physical differential rotation rate
, as it has a larger $\Omega_{\rm c}$ and smaller $\Omega_{\rm eq}$. 

\begin{figure}
	\begin{center}
		\includegraphics[width=0.95\columnwidth]{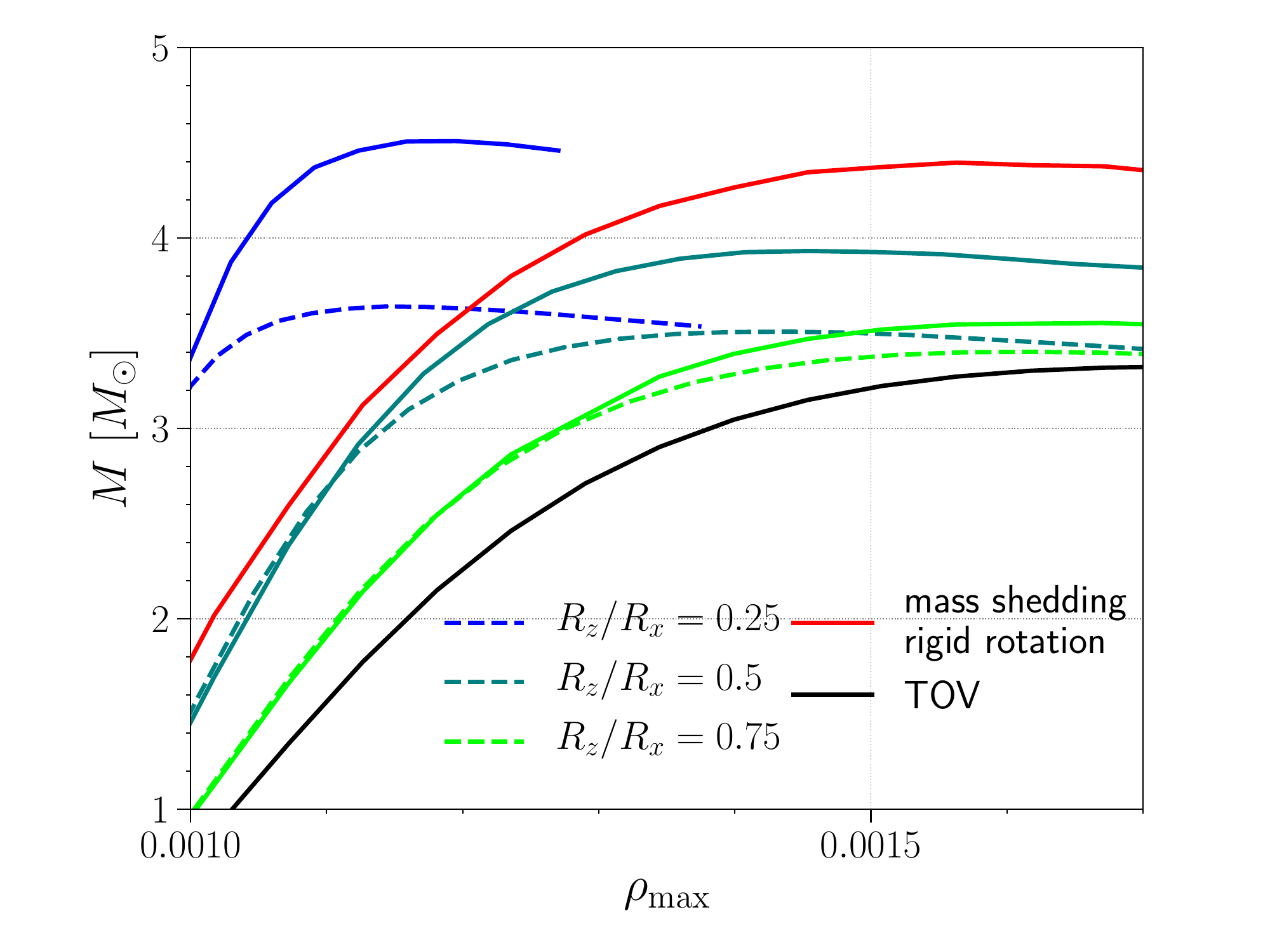}
	\end{center}
	\caption{Mass versus maximum density diagram for strangeon star model.
The black and red curves are for non-rotating and uniformly rotating mass
shedding limit case, respectively. The other curves ranging from green to blue colors
are differentially rotating solutions with constant axis ratio
$R_z/R_x=0.25,\,0.5$ and 0.75. The dashed curves are for models with $j$-const law
with $\hat{A}=1.0$ and solid curves are for models with the new rotation law
Eq.(\ref{eq:oj}). }
	\label{fig:lxoj}
\end{figure}

In order to probe the behavior above under the more realistic differential
rotation law Eq. (\ref{eq:oj}), we construct sequences of differentially
rotating stars with deformations $R_z/R_x=0.25,\ 0.5,\ 0.75$ in Fig.
\ref{fig:lxoj}.  The parameters are chosen such that
$\Omega_\mathrm{m}/\Omega_\mathrm{c}=1.1$ and
$\Omega_\mathrm{eq}/\Omega_\mathrm{c}=0.5$.  Both the $j$-const law (dashed
lines) and the new differential rotation law (solid lines) are shown for
comparison.  As it can be seen, with the new differential law the maximum mass
is increased compared with the $j$-const law case. The smaller the axis ratio is
(in other words, the faster the rotation), the more significant the difference
between the two cases. In the case of $R_z/R_x=0.25$, the maximum mass exceeds
the mass shedding limit for rigid rotation. 

However, as can be seen by
comparing the 'DR-LX-I' and 'DR-LX-II' models in Table.\ref{tab:solutions}, the
angular momentum and kinetic energy are also increased in the case of the
non-monotonic differential rotation law as a trade off for a higher maximum
mass.  The angular momentum and kinetic energy of the merger remnant originate
from the binary inspiral stage, which should be independent of the rotation law.
Hence, for merger events, only comparing the remnant mass to the mass shedding
limit might not be sufficient enough to tell the real outcome of the merger
product, especially for the case that the remnant normally wouldn't obtain enough angular
momentum to reach the mass shedding limit.  In this case, investigating the
relationship between the maximum mass for a given angular momentum will be
particularly useful, which we will explore in the next subsection.

\subsection{Critical mass of constant angular momentum sequences}
\label{sec:mcri}

One of the most important results in the theory of stability of rigidly rotating stars
is the ``turning point'' theorem of Friedman, Ipser, and Sorkin
\citep{Friedman88} which states that along a sequence with a constant angular
momentum $J$ and varying mass and central density, secular instability sets in
at the maximum mass i.e. at the turning point of the $M-\rho$ curve. 
The conjecture that similar to uniformly rotating stars, the dynamical
stability line also exists in differentially rotating stars was proven in the
affirmative at \cite{Weih2017} and thus the turning-point criterion can be used
as a first approximation for finding the critical mass for prompt collapse to a
black hole. We refer to this critical mass by $M_\mathrm{crit}$ hereafter \footnote{
In practice, we find $M_{\rm crit}$ by finding the point where $\frac{\partial M}{\partial \rho_{\rm max}}|_J=0$}.

Inspired by the fact mentioned in the previous subsection, that the maximum mass
of HMSS correlates with its angular momentum, it is interesting to investigate
whether HMSSs follow a similar universal relationship revealed by
\citep{Bozzola2018}. In particular, it has been found that the relationship
between $M_\mathrm{crit}$ and $J$ is $\hat{A}$-insensitive. Furthermore, when
re-normalized by the TOV maximum mass, the relationship between dimensionless
critical mass and angular momentum is found to be independent on EoSs of NSs
\citep{Bozzola2018}. In other words, for any NS EoSs, the enhancement in maximum
mass is determined only by the angular momentum of the rotating star but not how
the angular momentum is distributed inside the star. The reason that a HMNS can
have a larger maximum mass than a SMNS is because a HMNS can reach larger
angular momentum. Although in \citep{Bozzola2018} it has been shown that this
EoS-independent relationship cannot be extended for the case of even uniformly
rotating SSs, it's quite useful if one can at least verify whether the
$\hat{A}$-insensitive relationship still holds for differentially rotating SSs.

\begin{figure}
	\begin{center}
		\includegraphics[width=0.95\columnwidth]{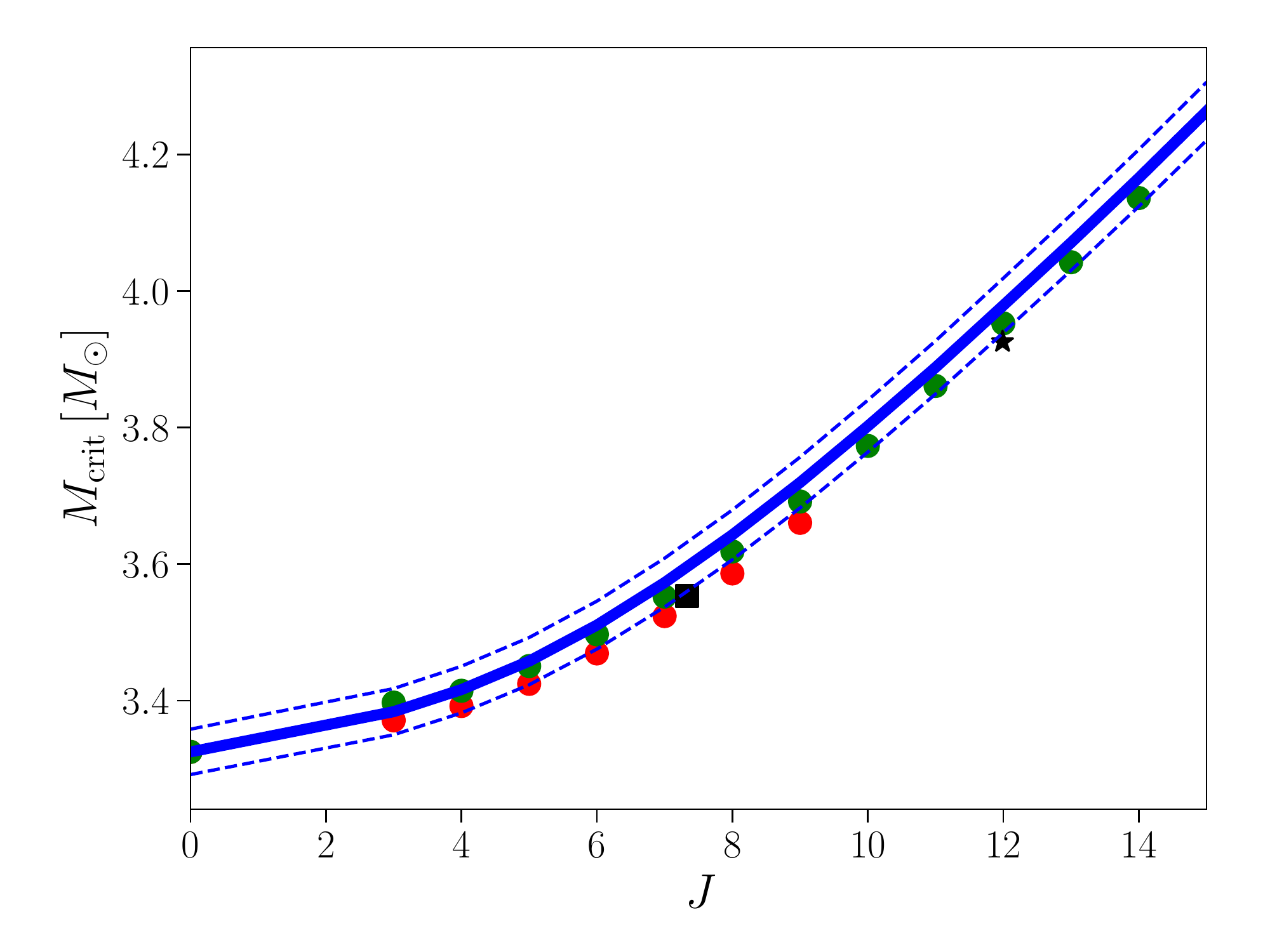}
		\includegraphics[width=0.95\columnwidth]{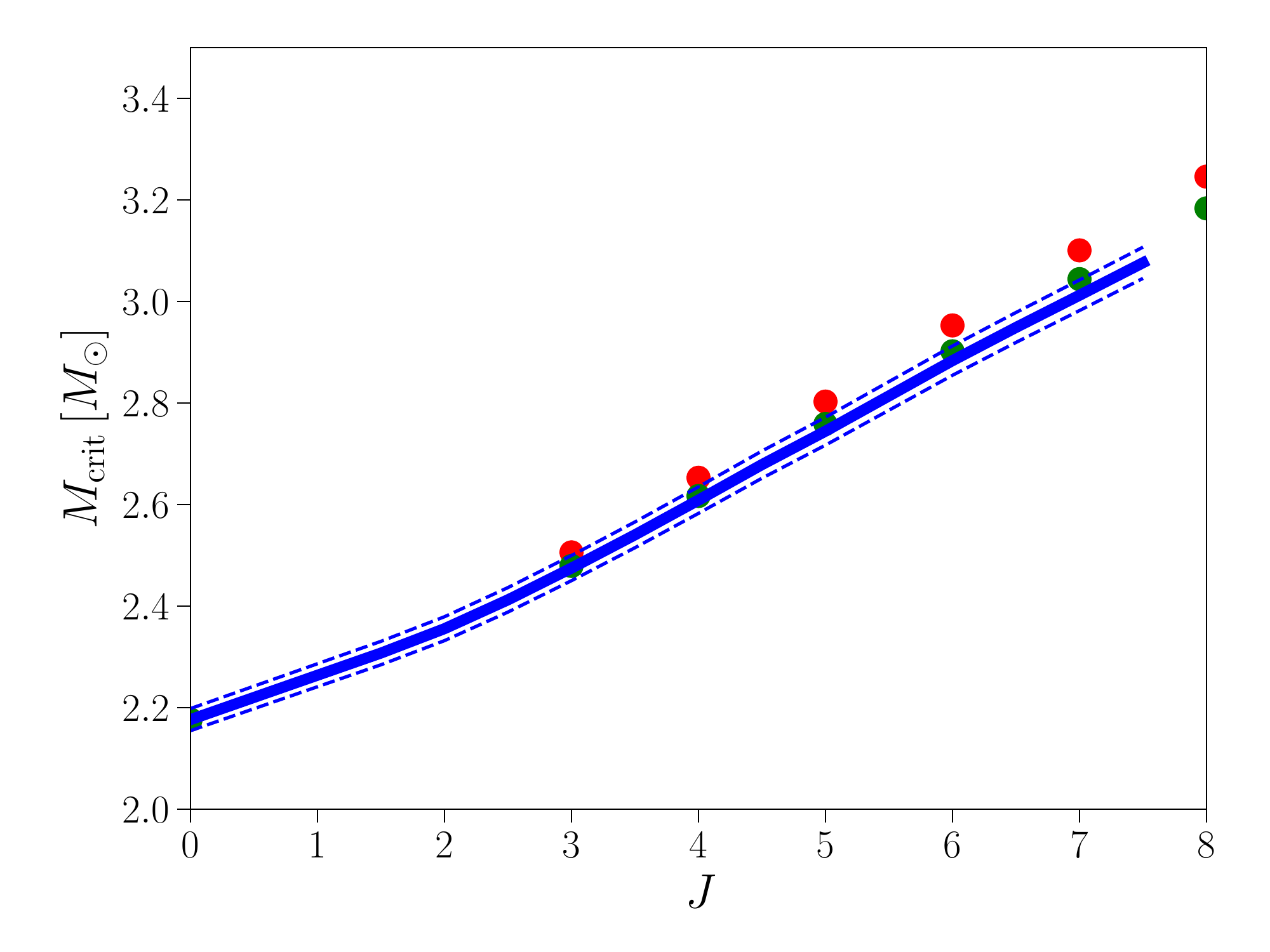}
	\end{center}
	\caption{The relationship between critical mass $M_\mathrm{crit}$ and
angular momentum $J$ for strangeon stars (upper panel) and MIT bag model stars
(lower panel). Both rigid rotating case (solid blue line) and differentially
rotating case (green dots for $\hat{A}=3.0$ and red dots for $\hat{A}=1.0$) are
shown. The $1\%$ error range for the relationship of the rigid rotating case
is shown in dashed blue lines for comparison purpose. As can be seen, for both
EoSs even in the case of $\hat{A}=1.0$, the relationship between
$M_\mathrm{crit}$ and $J$ is still reasonable consistent with the rigid
rotating case. In the upper panel, we have also labeled the results from the new
differential rotation law with black markers.}
	\label{fig:diff}
\end{figure}

We have considered the case of $\hat{A}=1.0$ and 3.0 for both strangeon star
model and MIT bag model to test the relationship between $M_\mathrm{crit}$ and
$J$. The results are shown in Fig. \ref{fig:diff}, where the rigid rotation case
(solid blue line) and the differential rotation case (colored dots) are
compared. As can be seen, even though $\hat{A}=1.0$ already represents a large
differention rotation degree for SSs, the $M_\mathrm{crit}$-$J$ relation
doesn't deviate much from the rigid rotation case (which is
$\hat{A}\rightarrow\infty$) for both EoSs. The relative difference as defined in
\citep{Bozzola2018} satisfies
\beq
\frac{f_\mathrm{uni}-f_{\hat{A}}}{f_\mathrm{uni}}\le2.0\%\quad\forall\hat{A}>1.0,
\label{eq:diff}
\eeq
for SSs too, where $f_\mathrm{uni}$ denotes $M_\mathrm{crit}$ for a certain $J$
for uniform rotation case and $f_{\hat{A}}$ for differential rotation case.
According to the upper panel in Fig.\ref{fig:diff}, the angular momentum of a
differentially rotating strangeon star can reach is much smaller than that of
the rigid rotating case. This explains why a HMSS could have a smaller maximum
mass than SMSS. What's more interesting is that, as can be seen from the upper
panel of Fig.\ref{fig:diff}, the solutions with the new differential rotation
law are also found to follow this relation between $M_\mathrm{crit}$ and $J$.
This result excludes the possibility that this relationship is due to a choice
of any particular differential rotation law. Hence, \textit{one can try to infer
the outcome of a binary merger event without having to know the details of the
rotational profile in the merger remnant}.

%

\begin{figure}
  \begin{center}
    \includegraphics[width=0.95\columnwidth]{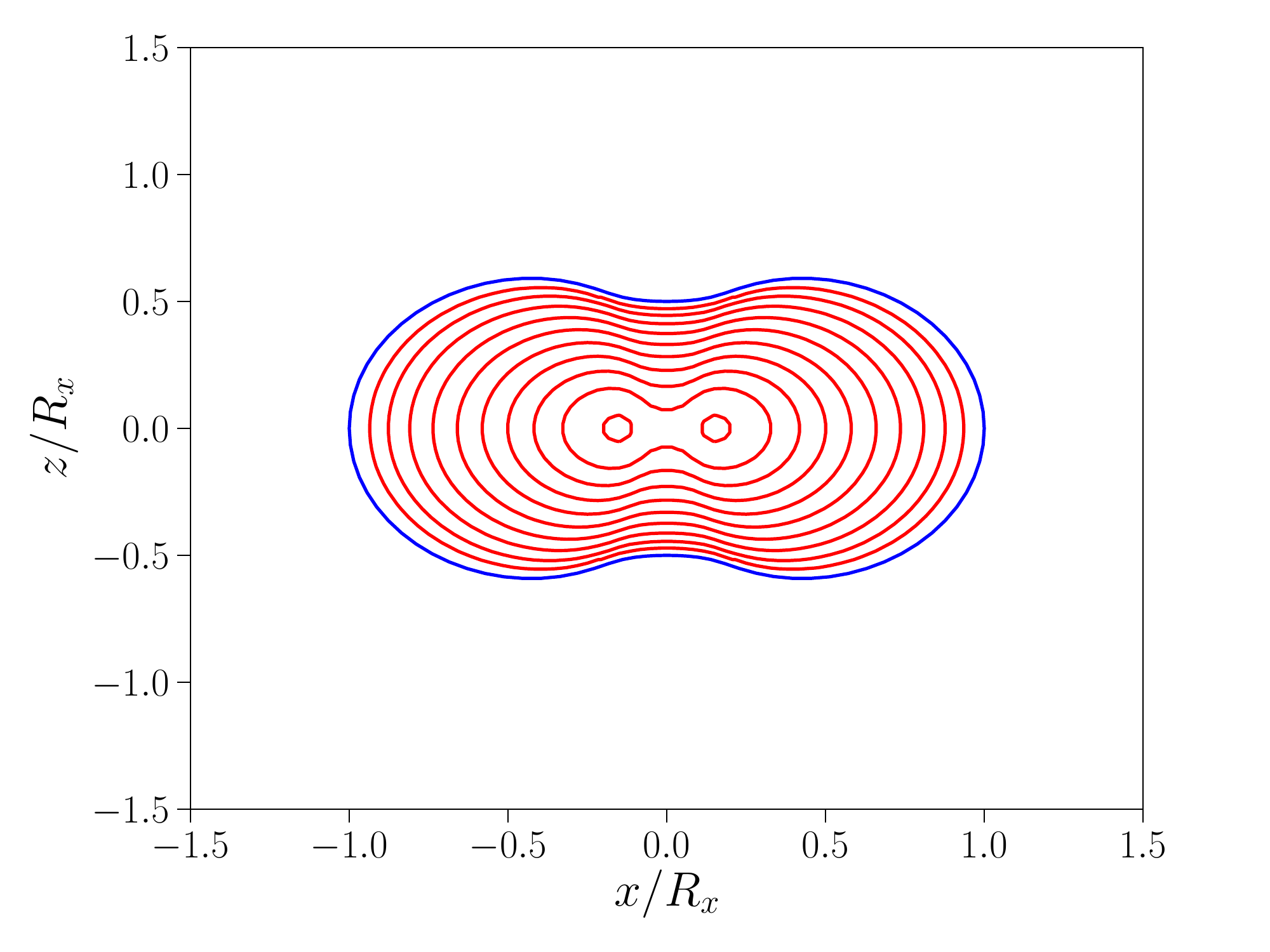}
    \includegraphics[width=0.95\columnwidth]{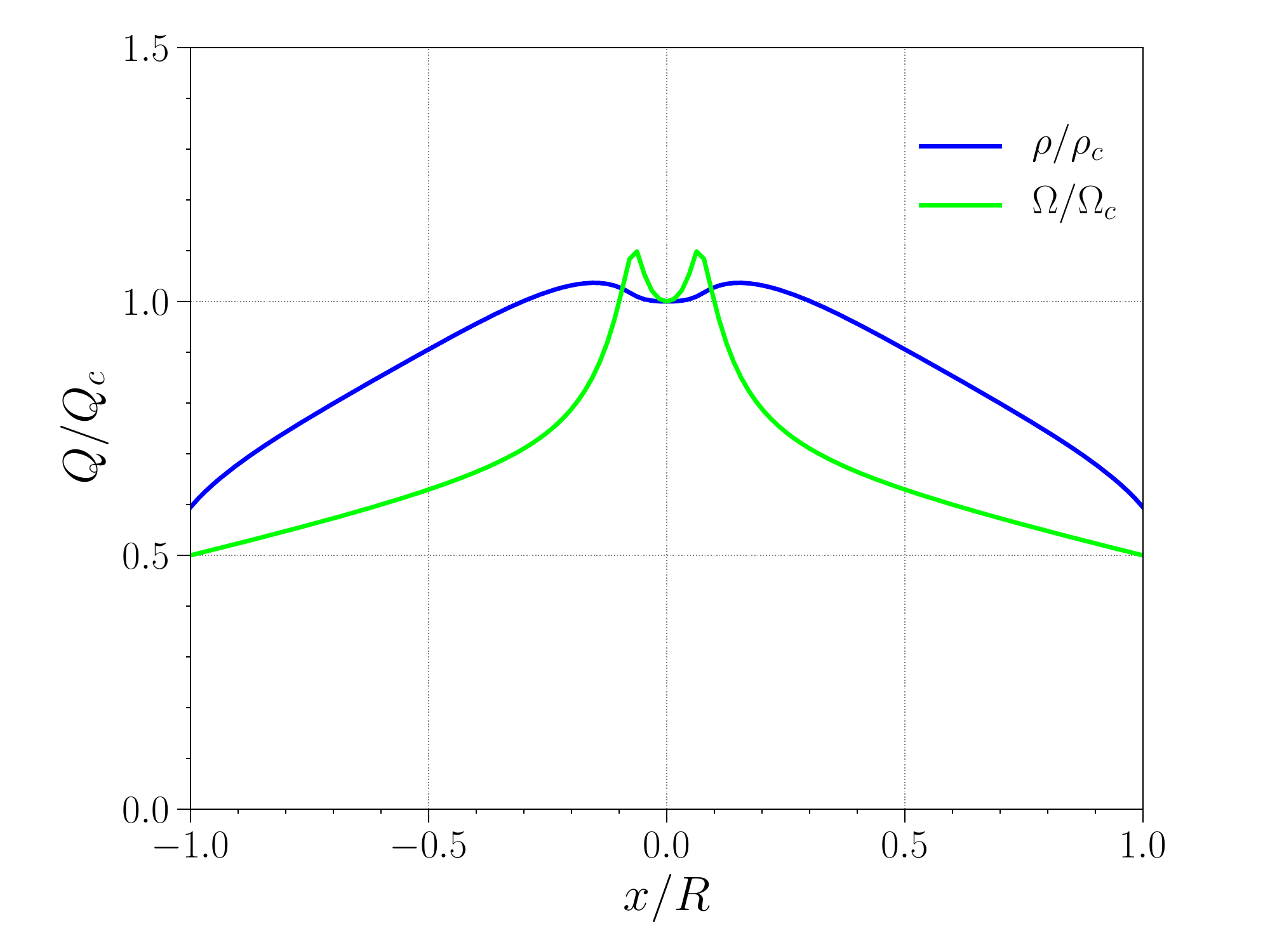}
  \end{center}
  \caption{Stellar surface and rest mass density contour of a differentially
rotating strangeon star (upper panel) and its density and angular velocity
profile (lower panel). Details about the solution shown in this figure can be
found in the 'DR-LX-4' model in Table.\ref{tab:solutions}}
	\label{fig:lxcontour}
\end{figure}

\subsection{Type C solutions of differentially rotating strange star}
\label{sec:typec}

Another interesting and important feature of differentially rotating
relativistic stars is the existence of different types of solutions according to
their geometrical surface shape, namely spheroidal or toroidal classes
\citep{Ansorg2009}. By using \cocal, we are able to construct and study the Type
C solutions of differentially rotating SSs according to the classification in
\citep{Ansorg2009}.  For rigidly rotating relativistic stars or differentially
rotating stars with relatively weak differential rotation rates, the solution
sequences terminate at the so-called mass-shedding limit with a finite axis
ratio $R_z/R_x$. Nevertheless, with a relatively strong differential rotation
degree, the solution sequence could go through a continuous transition to a
toroidal class with $R_z/R_x=0$. In such solution sequences, the stellar surface
in the $x-z$ plane may look like a peanut-shape and the maximum density is no
longer in the center of the star but in a ring of a finite radius inside the
star (c.f. Fig. \ref{fig:lxcontour} as an example). Identifying such solutions
for differentially rotating SSs is helpful in determining the maximum mass as
well as in understanding the influence of a certain differential rotation rate. 

According to the parameter study for the solution space of differentially
rotating NSs \citep{Rosinska2017}, type C solutions come to exist for
$\hat{A}\lesssim1.0$ \footnote{Note that in \citep{Rosinska2017} the definition
of $\tilde{A}$ is different from $\hat{A}$ used in this paper, but are related
simply as $\tilde{A}=1/\hat{A}$.}, although a more precise value depends on the
central density. In order to make a comparison we have also tested the $j$-const
law for SSs.  Properties of selected type C solutions for differentially
rotating SSs are listed in Tab. \ref{tab:solutions}. It turns out that type C
solutions emerge at much larger $\hat{A}$, thus much smaller differential
rotation rate. For instance, for both strangeon star and MIT bag model with
$\hat{A}=3.0$ (which corresponds to $\tilde{A}=1/3$ in Fig. 5 in
\citep{Rosinska2017}), toroidal solutions are already found for the whole
central density range we considered. 

We have identified the first solution in a sequence, the maximum density of
which is no longer at the center of the star, as the beginning of the transition
to the toroidal class\footnote{Identically, one can also try to find the first
solution, the surface of which in the $x-z$ plane is no longer elliptical.}. By
doing so, we realize that the transition happens at an axis ratio very close to
1 for differentially rotating SSs with $\hat{A}=1$. In other words, with very
little angular momentum, the differential rotation is already playing an
important role in the changing of the configuration of a SS. One such solution
is also listed as 'DR-LX-3' in Tab. \ref{tab:solutions} to illustrate the onset
of this transition.

Similar analysis has been conducted for the solutions with the new differential
rotation law. Although as mentioned above, it's not easy to have a solution with
very small axis ratio as it's increasingly difficult to adopt the $A$ and $B$
parameter for smaller axis ratios. Despite of that, we still managed to reach
$R_z/R_x=0$ and find toroidal solutions for the low central density sequence for
the case used in our calculation ($\Omega_\mathrm{m}/\Omega_\mathrm{c}=1.1$ and
$\Omega_\mathrm{eq}/\Omega_\mathrm{c}=0.5$). For relatively large central
density sequence, we attempt to figure out whether the transition to toroidal
class is already triggered by looking at the stellar surface and density profile
of the star. The result shows that for $R_z/R_x=0.5$ case, the onset of the
transition already happens for all the central density range (an example can be
found in Fig.\ref{fig:lxcontour}). Hence, this type C solution should be a
common feature for differentially rotating relativistic stars, regardless of the
EoSs and details of the rotation profile.

\begin{table*}
		\begin{tabular}{cccccccccccc}
		\hline
		Model & $R_x$ & $R_z/R_x$ & $\rho_c$ & $\rho_{\rm max}$ & $\Omega_c$ & $M_\mathrm{ADM}$ & $J$ & $T/|W|$
		\\
		\hline
		UR-LX    & 4.82 (15.1) & 0.53125 (0.584)   & $1.56\times10^{-3}$ & $1.56\times10^{-3}$ & 0.0603 & 4.39 & 16.4 &	0.222 	 \\
		DR-LX-1  & 4.36 (12.4) & 0.015625 (0.0190) & $8.68\times10^{-4}$ & $1.51\times10^{-3}$ & 0.382  & 3.78 & 10.3 & 0.183   \\
		DR-LX-2  & 4.07 (14.4) & 0.25 (0.295)      & $1.20\times10^{-3}$ & $1.40\times10^{-3}$ & 0.110  & 4.49 & 17.6 & 0.290    \\
		DR-LX-3  & 4.83 (10.9) & 0.9375 (0.947)    & $1.51\times10^{-3}$&  $1.51\times10^{-3}$ & 0.0638 & 3.25 & 2.28 & 0.0135   \\
		DR-LX-4  & 4.26 (12.8) & 0.50 (0.553)      & $1.46\times10^{-3}$ & $1.51\times10^{-3}$ & 0.0945 & 3.92 & 11.9 & 0.203    \\
		UR-MIT   & 8.23 (15.1) & 0.484375 (0.523)  & $1.76\times10^{-3}$ & $1.76\times10^{-3}$ & 0.0433 & 3.17 & 8.56 & 0.198  	 \\
		DR-MIT-1 & 6.79 (13.9) & 0.015625 (0.0172) & $6.07\times10^{-3}$ & $1.34\times10^{-3}$ & 0.163  & 3.60 & 10.8 & 0.236 	 \\
		\hline
	\end{tabular}
	\caption{Quantities of selected solutions for rotating SSs. In the
above, $R_x$ is the coordinate (proper) equatorial radius and $R_z/R_x$ is the
ratio of coordinate (proper) polar to the equatorial radius. $\rho_c$ is the
central rest-mass density and $\rho_{\rm max}$ the maximum rest-mass density in
the star. $\Omega_c$, $M_{\rm ADM}$, $J$, and $T/|W|$ are the central angular
velocity, Arnowit-Deset-Misner mass, angular momentum and ratio between kinetic
energy and gravitational potential.  Definitions can be found in the Appendix of
\citep{Uryu2016a}. In this table, 'UR-LX' and 'UR-MIT' labels the maximum mass
solution of uniformly rotating strangeon star and MIT bag model star,
respectively. 'DR-LX-1' and 'DR-MIT-1' are the maximum mass solutions for
differentially rotating strangeon star and MIT bag model star with $\hat{A}=1$
$j-$const law. 'DR-LX-2' is the maximum mass solution for the new differential
rotation law with $R_z/R_x=0.25$ for the strangeon star model. 'DR-LX-3' and
'DR-LX-4' are two selected type-C solutions with $j-$ const law and the new
differential rotation law, respectively. } 
\label{tab:solutions}
\end{table*}

\section{Discussion and Conclusion}
\label{sec:disandconclu}

In this paper, we have calculated differentially rotating SSs, with both MIT bag
model and strangeon star model. Besides the widely used $j$-const law, we
have also considered a more realistic non-monotonic rotation profile. The
maximum mass of HMSSs,  toroidal solutions and the relationship between the
critical mass and angular momentum are investigated and compared with previous
results of HMNSs. Two major differences are found between HMNSs and HMSSs:
first, with a moderate differential rotation rate, the maximum mass of a HMNS
is increased significantly as the $\hat{A}$ parameter decreases (from 2.0 to
1.0). Whereas in the same range, the maximum mass of a HMSS drops significantly.
In particular, the maximum mass drops below the
rigid rotation case with a moderate differential rotation rate.
Secondly, the continuous transition to the toroidal
solutions happens at much larger $\hat{A}$, i.e. much smaller differential
rotation rate (typically $\hat{A}=3.0$ compared with $\hat{A}=1.0$ in the case of NSs). Both
differences indicate that a moderate differential rotation degree
for NSs is already too large for SSs. The self-bound nature of SSs can account
for this difference, as a certain difference in the angular velocity will play a
more important role for SSs, the density of which is almost uniform inside the
star. 
Despite these differences, similarly to NSs, a universal relationship between
$M_\mathrm{crit}$ and $J$ is found for SSs, even for the new differential 
rotation law. This provides a more realistic way to interpret the 
outcome of a binary merger event, rather than compare the remnant mass with the 
maximum mass. 

Combining all the results we have obtained in this paper, one conclusion we can draw
on the differentially rotating SS remnant formed in a binary merger event is that it's most 
likely to be a type C solution whose maximum density is not at the center. Meanwhile, 
due to the self-bound nature, the moment of inertia of SSs is larger than NSs and hence the 
$T/|W|$ ratio (similar results have already been reported in \citep{Zhou2018} and the resulting
secular instability for uniformly rotating SSs are studied).
According to previous studies
on the dynamical instabilities \citep{Centrella:2001xp,Saijo2003,Shibata:2002mr,Shibata:2003yj} 
of differentially rotating NSs, for the extremely
differential rotation rate cases (especially for the case the maximum density is no longer in the center, 
c.f. the discussions in \citep{Saijo2003}), 
the $T/|W|$ ratio for onset of such dynamical instabilities could be
reduced significantly. Consequently, such instabilities may easily take place if a 
differentially rotating SS is formed in a binary merger, redistributing matter
and angular momentum inside the star and destroying the toroidal shape of the star
in a few central rotation periods,
thus producing additional signatures in the GW radiation of the post-merger phase. At the same time,
such instability will compete against other mechanism such as magnetorotational instability in dissipating 
the differential rotation, whereas the later one is known to be responsible to enhance magnetic field 
of the merger remnant with the differential rotational kinetic energy.
Therefore, the remnant SS 
might have significantly smaller dipole magnetic fields compared with a NS remnant scenario,
providing a way to distinguish between a BSS and BNS merger scenario with the EM counterparts.

\acknowledgments 
E.Z. would like to thank Luciano Rezzolla for his warm host in the Relastro group in 
Uni Frankfurt and for useful discussions with the group members. This work was supported
by the National Key R\&D Program of China (Grant No. 2017YFA0402602), the National Natural
Science Foundation of China, and the Strategic Priority Research Program of Chinese Academy 
Sciences (Grant No. XDB23010200).   
A.T. was supported by NSF Grants No. PHY-1602536 and No. PHY-1662211 and NASA
grant 80NSSC17K0070 to the University of Illinois at Urbana-Champaign. K.U.  was
supported by JSPS Grant-in-Aid for Scientific Research (C) 15K05085 and 18K03624
to the University of Ryukyus. M.S. was supported by JSPS Grant-in-Aid 
for Scientific Research (A) 16H02183. The simulations were performed on the clusters 
LOEWE (CSC, Frankfurt) and Yoichi (AEI, Potsdam).

\bibliographystyle{apsrev4-1}
\bibliography{aeireferences}

\end{document}